\documentclass[twocolumn,amsmath,prl,aps,superscriptaddress,showpacs,a4paper]{revtex4}

\usepackage{graphicx}
\usepackage{placeins}

\begin{document}

\title{Hydrodynamic and Brownian Fluctuations in Sedimenting Suspensions}
\author{J.\ T.\ Padding}
\affiliation{Department of Chemistry, Cambridge University,
           Lensfield Road, Cambridge CB2 1EW, UK}
\affiliation{Schlumberger Cambridge Research, High Cross,
           Madingley Road, Cambridge CB3 0EL, UK}
\author{A.\ A.\ Louis}
\affiliation{Department of Chemistry, Cambridge University,
           Lensfield Road, Cambridge CB2 1EW, UK}
\date{\today}

\begin{abstract}
We use a mesoscopic computer simulation method to study the interplay
between hydrodynamic and Brownian fluctuations during  steady-state
sedimentation of hard sphere particles for Peclet numbers (Pe) ranging
from $0.1$ to $15$.  Even when the hydrodynamic interactions are an
order of magnitude weaker than Brownian forces, they still induce
backflow effects that dominate the reduction of the average
sedimentation velocity with increasing particle packing fraction.
Velocity fluctuations, on the other hand, begin to show nonequilibrium
hydrodynamic 
character for Pe > $1$.
\end{abstract}


\maketitle

Exactly what happens when a collection of particles sediments through
a viscous solvent is a very simple question to pose, but a remarkably
difficult one to answer\cite{Ramaswamy,Russ89}.  In a classic tour de
force, Batchelor\cite{Batchelor} showed that the average sedimentation
velocity $v_{s}$ of hard spheres (HS) of hydrodynamic radius $a$ has a
lowest order correction  $v_{s} = v_{s}^0(1-6.55 \phi)$ where $\phi$
is the HS volume fraction and $v_s^0$ is the Stokes sedimentation
velocity of a single sphere\cite{Russ89}.  This substantial correction
to $v_s^0$ with volume fraction is caused by many-body hydrodynamic
backflow effects that greatly complicate efforts to extend the
Bachelor result to higher order in $\phi$.  Even less is understood
about the velocity fluctuations around the average, $\delta v =
v-v_s$.  Using straightforward physical arguments, Caflish and
Luke\cite{CaflishLuke} predicted that for sedimentation these should
diverge as $\left \langle(\delta v)^2\right\rangle \sim L$, where $L$
is the smallest container size.  This surprising result provoked a
flurry of experimental and theoretical studies (see
ref.\cite{Ramaswamy} for a review).  Although it is agreed that
hydrodynamic velocity fluctuations are relatively large, there is no
consensus on the reasons (if any) for the purported breakdown of the
Caflish-Luke argument at large~$L$.

 In addition to the fundamental interest and myriad applications of
 sedimentation itself, researchers have been motivated to investigate
 this problem because of its relevance to non-equilibrium statistical
 mechanics. Recent studies in this line include a theoretical
 prediction of a continuous nonequilibrium noise-driven phase
 transition between screened and unscreened
 phases\cite{Levi98}, and an experimental study predicting a
 noise-induced effective ``temperature'' that  could aid in
 developing an ensemble based statistical mechanics for driven
 systems\cite{Segr01}.

Most  theoretical studies of sedimentation have focused on
the limit where Brownian forces are negligible, and only hydrodynamic
interactions (HI) contribute. In other words, the dimensionless
Peclet number
\begin{equation}
\mathrm{Pe} = \frac{v_s^0 a}{D_{col}},
\label{defPe}
\end{equation}
which measures the relative strength of HI and Brownian forces, was
assumed to be infinite. Here $D_{col}$ is the equilibrium
self-diffusion constant of the particles.
 When the gravitational energy gained by a particle sedimenting over a
 distance of one radius $a$ is equal to the reduced temperature
 $k_BT$, then $\mathrm{Pe}=1$\cite{Ramaswamy}, a criterion  used to define the
 start of the colloidal regime\cite{Russ89}.  Sedimentation at $\mathrm{Pe}
 \leq 1$ has many important applications for colloidal
 dispersions, as well as for centrifugal diagnostic techniques
 commonly used for biological macromolecules\cite{Russ89}.

In this Letter, we employ a recently proposed mesoscopic simulation
method \cite{Male99} to investigate steady state
sedimentation at finite Pe, where Brownian and HI both
contribute to velocity fluctuations.  To our knowledge, this problem
has not been investigated in detail before. For all Pe studied,
we find that the average sedimentation velocity is completely
dominated by HI, even when  they are much smaller than the
Brownian forces.  On the other hand, we argue that short time velocity
fluctuations are dominated by Brownian forces up to surprisingly large
Pe, while long time fluctuations have predominantly hydrodynamic
character even at moderate Pe.

To perform the simulations, we adapt Stochastic Rotation Dynamics (SRD)
\cite{Male99} to the problem 
of sedimenting HS.  SRD is a particle based method similar in spirit
to the lattice Boltzmann model (LB), which has been extensively
applied to sedimentation\cite{Ladd1996}. In contrast to LB, it
naturally includes Brownian noise (see however
\cite{Cates}).  In
SRD a fluid is represented by $N_f$ ideal particles of mass $m_f$.  After
propagating the particles for a time $\Delta t_c$, the system is
partitioned into cubic cells of volume $a_0^3$.  The velocities
relative to the center of mass  velocity of each separate cell are
 rotated over a fixed angle around a random
axis. This procedure conserves mass, momentum, and energy, and yields
the correct hydrodynamic (Navier Stokes) equations,
\textit{including} the effect of thermal noise
\cite{Male99}. 
The fluid particles only interact with each other through the rotation
procedure, which can be viewed as a coarse-graining of particle
collisions over time and space.  For this reason, the particles should
not be interpreted as individual molecules but rather as a Navier Stokes
solver that naturally includes Brownian noise.

The colloid-colloid (cc) and colloid-fluid (cf) interactions are modeled
by a repulsive potential:
 $\beta V_{ci}(r) = 10
\left[ (\sigma_{ci}/r)^{2n} - (\sigma_{ci}/r)^n + 1/4 \right]$ ($r
\leq 2^{1/n}\sigma_{ci}$).  For $V_{cc}(r)$,  $n=24$, which should  approximate HS behavior.
For $V_{cf}(r)$ we set $n=6$ and $\sigma_{cf} = 0.465 \sigma_{cc}$,
slightly below half the colloid diameter $\sigma_{cc}$, which allows for
lubrication.  These potentials result in a hydrodynamic radius $a
\approx 0.8 \sigma_{cf}$.  Colloid-colloid and colloid-fluid forces are 
integrated with 
a standard velocity Verlet molecular
dynamics integrator with a time step $t=\frac14 \Delta
t_c$\cite{noslip}.

We now briefly discuss our choice of SRD parameters; a more detailed
account will be published elsewhere\cite{PaddingLouis}.  The kinematic
viscosity $\nu = \eta_f/\rho_f$, where $\eta_f$ is the viscosity and
$\rho_f$ the mass density of the fluid, is an important parameter
because it sets the timescale over which the momentum (vorticity)
diffuses away. In dimensionless form it is desirable to have
$\nu/D_{col} > Sc=\nu/D_f \gg1$, where $Sc$ is the Schmidt number and
$D_f$ the self-diffusion constant of the fluid particles. Since
$D_{col} < D_f$, the first inequality is always satisfied.  When $Sc
\approx 1$ momentum diffusion is dominated by mass diffusion, as in a
gas. If $Sc \gg 1$ the fluid is liquid like.  Since the SRD particles
could be viewed as collections of individual molecules, the Schmidt
number of an SRD fluid is smaller than the real fluid it
represents.  Nevertheless, to model a liquid-like system, we choose a
relatively small collision interval ($\Delta t_c = 0.1 a_0
(m_f/k_BT)^{1/2}$), leading to $Sc \approx 5$. To prevent
compressibility effects, the gravitational field $g$ was limited such
that $0.0067 \leq v_s/c_f \leq 0.1$, where $c_f=\sqrt{2 k_B T/m_f}$ is
the speed of sound in the fluid.  Finally, to avoid large inertial
effects, the particle Reynolds number $Re = v_sa/\nu$ should be $\ll 1$,
as in real suspensions\cite{Russ89}.  Inevitably there will be a
compromise between computational efficiency and low $Re$.  In our work
$0.0016
\leq Re \leq 0.24$, depending on Pe, which is  similar to the
choice made for LB simulations\cite{Ladd1996,Cates}.

To further test the accuracy of our method, we measured the Stokes
drag $F_d$ on the colloid for various values of the sphere radius
$\sigma_{cs}$ and gravitational field $g$.  By varying the box size we
find excellent agreement with  analytic finite-size
corrections\cite{Zick82}, from which the infinite box-size limit
extrapolates to $F_d = \gamma v_s = 4\pi
\eta_f a v_s$, as expected for slip boundary conditions\cite{noslip}.  For the
largest box sizes we compared the full velocity field around a single
colloid to the known analytic result\cite{Russ89}, and 
varied the ratio $a_0/a$, finding that errors scale roughly
linearly with this parameter.  We choose $a_0/a \approx \frac12$,
which leads to a relative error in the full velocity field of about
$2\%$, similar to what is used in LB\cite{Ladd1996,Cates}, and
sufficiently accurate for the kinds of questions we
investigate\cite{PaddingLouis}.

\begin{figure}[t]
  \scalebox{0.40}{\includegraphics{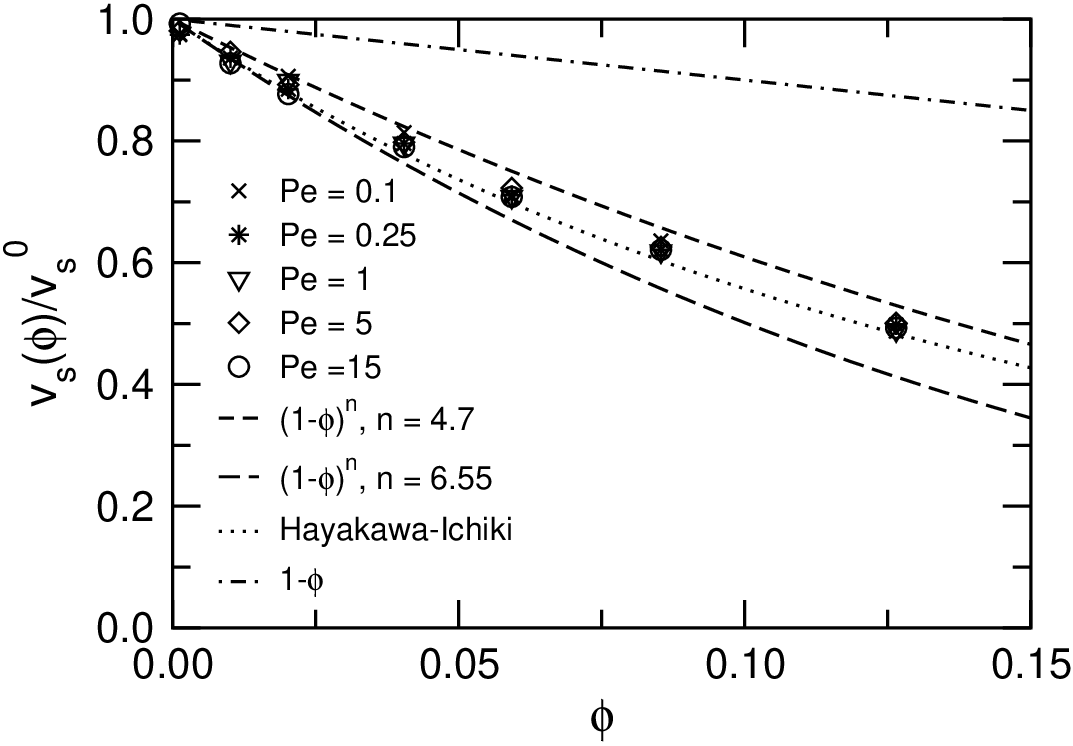}}
\caption{Average sedimentation velocity, $v_s$ normalized by the Stokes velocity
$v_s^0$, as a function of volume fraction $\phi$ for various Peclet
numbers. The reduction of $v_s$ due to hydrodynamic backflow effects is
independent of Pe.  Dashed lines correspond to two versions of the
semi-empirical Richardson-Zaki law $v_s/v_s^0 =
(1-\phi)^n$\protect\cite{Russ89}. The dotted line is another
theoretical prediction taking higher order HI into
account\protect\cite{Haya95}.  Ignoring hydrodynamics leads to 
$v_s/v_s^0=1-\phi$ (dash-dotted line).
\label{fig1}
}\vspace*{-0.5cm}
\end{figure}

The sedimentation runs were performed in a periodic box of dimensions
$L_x = L_y = 32a_0$ and $L_z = 96a_0$, with $N = 8$ to 800 colloids
and $N_f \approx 5 \times 10^5$ SRD particles, corresponding to an
average of about $5$ particles per coarse-graining cell volume
$a_0^3$.  The system size is similar to some successful LB
simulations\cite{Ladd1996}. A gravitational field $g$, applied in the
$z$ direction, was varied to produce different Pe. The simulations
were run from $200$ Stokes times $t_S$ (for $\mathrm{Pe}=0.1$) to $30,000$
$t_S$ (for $\mathrm{Pe}=15$), where $t_S= a/v_s$ is the time it takes a sphere
to sediment one particle radius. We verified that there was no drift
in averages after about $100$ $t_S$, so that the suspension is in
steady state. To check that our system is large enough, we performed
some runs for double the box size described above, as well as for
$a_0/a=\frac14$,  finding no significant changes in our
conclusions\cite{PaddingLouis}.

The average sedimentation velocity $v_s$ for different Peclet numbers
and different sphere packing fractions $\phi = \frac43 \pi \rho a^3$,
with $\rho$ the colloid number density, is shown in Fig.~\ref{fig1}.
  At  low densities the results are
consistent with the Batchelor  law\cite{Batchelor},
while at higher densities they compare well with a number of other 
forms also derived for the $\mathrm{Pe}\rightarrow\infty$ limit\cite{Russ89}.
 Although one might naively expect that the
effect of HI becomes weaker for $\mathrm{Pe}<1$, we observe that the results
for
\textit{all} Peclet numbers $0.1 \leq \mathrm{Pe} \leq 15$ lie exactly on the
same curve. Taking into account only Brownian fluctuations gives
$v_s = v_s^0(1-\phi)$ \cite{Russ89}, which heavily underestimates backflow effects.
This is strong evidence that purely hydrodynamic arguments are still
valid in an
\textit{average} sense at low Pe.

We next discuss velocity fluctuations around the average. 
In colloidal systems the
instantaneous velocity fluctuations $\delta \mathbf{v} =\mathbf{v} -\mathbf{v}_s$ are
dominated by thermal fluctuations, with a magnitude determined by
equipartition:
$\Delta v_{T}^2 = k_BT/m$.
Here $m$ is the mass of a colloid. To disentangle the
hydrodynamic fluctuations from thermal fluctuations, we describe
spatial and temporal correlations in the velocity fluctuations. The
spatial correlation of the $z$ component (parallel to the
sedimentation) of the velocity fluctuations can be defined as
\begin{equation}\label{czz}
C_z(\mathbf{r}) \equiv \left \langle \delta v_z(\mathbf{0}) \delta v_z(\mathbf{r}) \right\rangle\mbox{,}
\end{equation}
where $\left\langle \ldots \right\rangle$ represents a time average over 
many particles.
The distance vector $\mathbf{r}$ is taken perpendicular to sedimentation,
$C_z(x)$, or parallel to it, $C_z(z)$.
Similarly, the temporal correlation of the $z$ component of the
velocity fluctuations can be defined as
\begin{equation}\label{czt}
C_z(t) \equiv \left \langle \delta v_z(0) \delta v_z(t) \right\rangle\mbox{.}
\end{equation}


\begin{figure}[t]
  \scalebox{0.40}{\includegraphics{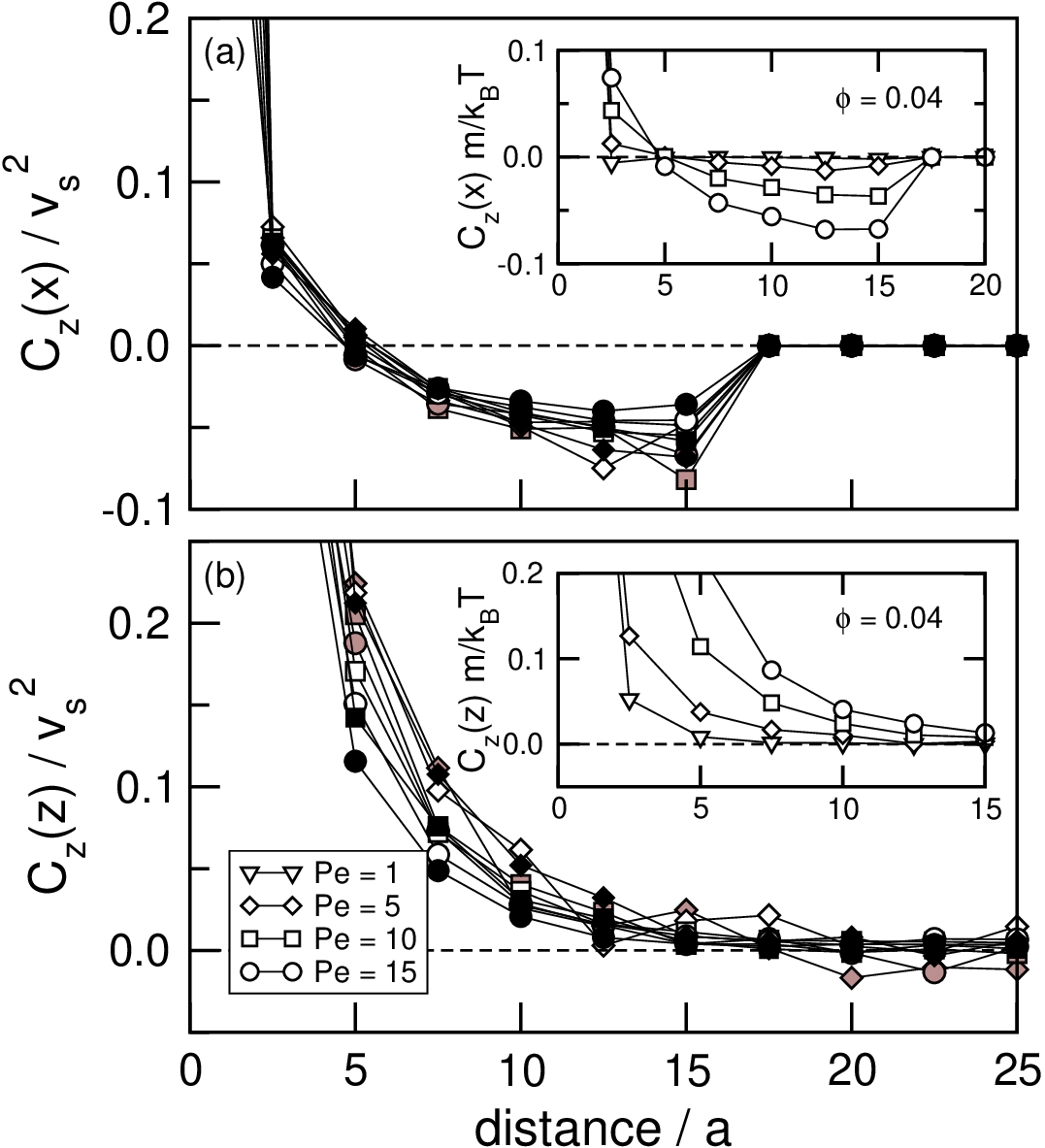}}
\caption{ Spatial correlation functions of the parallel ($z$) 
component of the velocity fluctuations as a function of distance
perpendicular (a) and parallel (b) to the external field, for three different
volume fractions ($\phi = 0.02$ (grey symbols), $\phi = 0.04$ (white),
$\phi = 0.086$ (black)) and different Peclet numbers. The correlation functions are scaled with $v_s^2$ to emphasize hydrodynamic fluctuations.
  The insets show how $C_{z}({\bf r})$, scaled  with
$C(0)=k_B T/m$, increases with Pe.
\label{fig2}
}\vspace*{-0.5cm}
\end{figure}

In Fig.~\ref{fig2} we plot $C_z({\bf r})$, which shows a positive
spatial correlation along the direction of flow, and an
anti-correlation perpendicular to the flow, very much like that
observed in experiments\cite{Nico95}.  The inset of Fig.~\ref{fig2}(a)
shows that at Pe=1 the correlation in the perpendicular direction,
$C_z(x)$, is almost negligible compared with the thermal fluctuation
strength $k_BT/m$, whereas for larger Pe, distinct regions of
negative amplitude emerge, which grow with increasing Pe. Similarly,
the inset of Fig.~\ref{fig2}(b) shows correlations in the parallel
direction that rapidly increase with Pe. For the highest Peclet
numbers studied ($5 \leq \mathrm{Pe} \leq 15$), the amplitudes of these
correlations grow proportionally to $v_s^2$, as shown in the main
plots of Fig.~\ref{fig2}. Unfortunately, because the division by
$v_s^2$ amplifies the statistical noise, we are unable to verify
whether this scaling persists for $\mathrm{Pe}< 5$. The minimum in
Fig.~\ref{fig2}(a) is limited by the box size.  We checked this by
simulating larger systems: the correlation size increased linearly
with box dimensions \cite{PaddingLouis}, as found for
LB\cite{Ladd1996}, suggesting that the hydrodynamic velocity fluctuations are unscreened.

The following timescales are important for temporal correlations: In a
liquid, the solvent relaxation time $\tau_f$, typically of order
$10^{-14}$s\cite{Russ89}, is the smallest relevant timescale.  A
(larger) Brownian particle experiences random forces and a friction
$\gamma$. As a consequence, it loses memory of its
initial velocity after a time $\tau_{B} \approx m/\gamma$, which is
typically of the order of $10^{-9}$ s\cite{Russ89}. For time scales
larger than $\tau_{B}$, the particle experiences diffusive
behavior and after a  time $\tau_D = a^2/D
\gg
\tau_{B}$ it has traveled over its own radius. 
For correct coarse-grained temporal behavior, the timescales don't
need to be identical to the underlying fluid, but it is important that
they are clearly separated\cite{Cates}. This is indeed the case for
our choice of SRD parameters, where $\tau_f \approx \Delta t_c = 0.1$,
$\tau_B
\approx 2$, and $\tau_D
\approx 200$ (in units $a_0(m_f/k_B T)^\frac12$). 
Since the Stokes time $t_S \equiv a/v_s
= \tau_D/\mathrm{Pe}$  must be $\gg
\tau_B$, this 
this sets a limit on the maximum Pe number for these parameters.

\begin{figure}[t]
  \scalebox{0.40}{\includegraphics{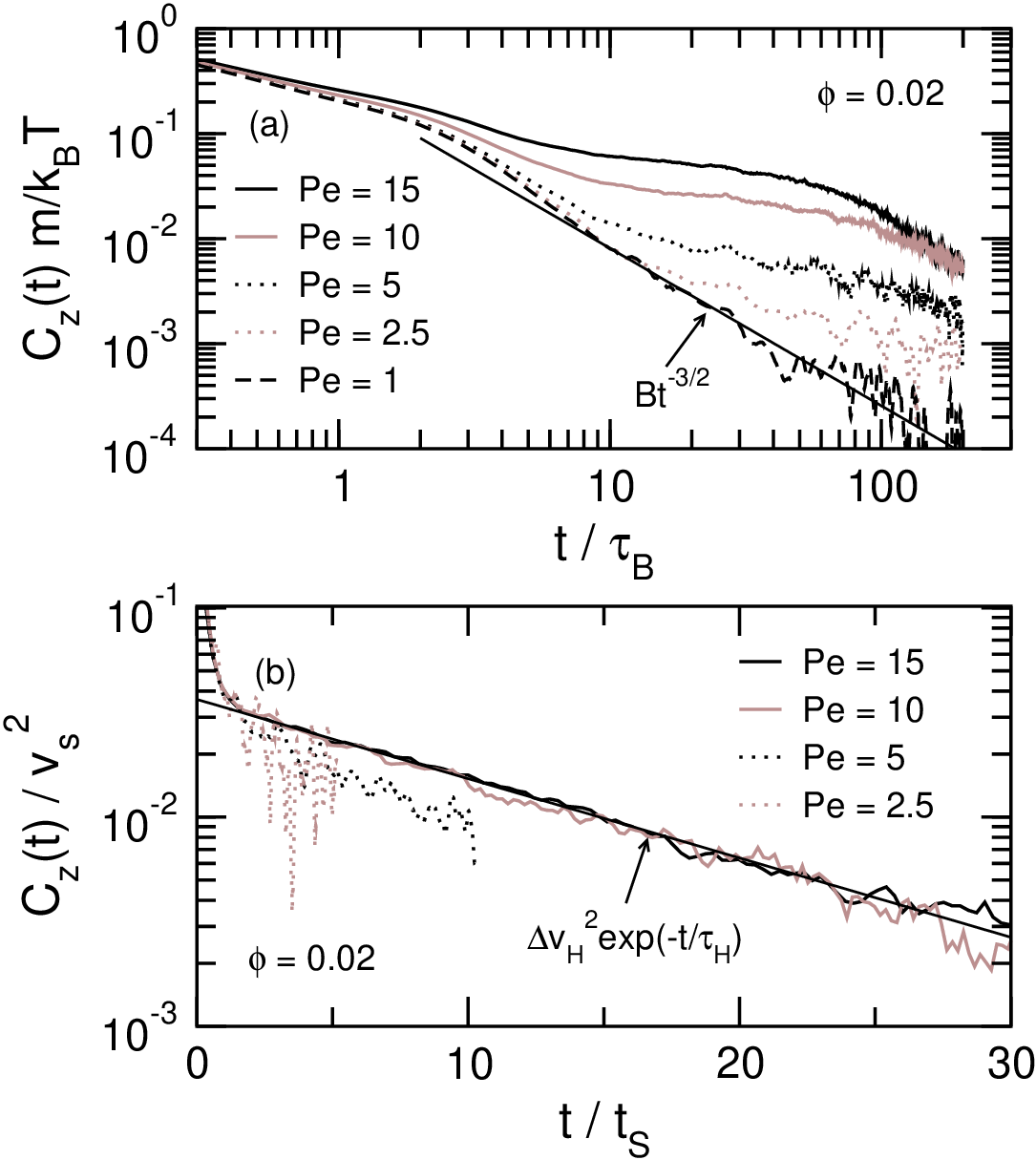}}
\caption{Temporal correlation functions of the $z$ component 
of the velocity fluctuations for $\phi = 0.02$ and different Peclet
numbers.  (a) Time is scaled with the Brownian relaxation time
$\tau_B=m/\gamma$ and the velocities are scaled with the thermal
fluctuation strength $k_BT/m$. The straight line is the hydrodynamic
long time tail $B t^{-3/2}$ with  $B^{-1} = 12\rho_fk_BT(\pi
\nu)^{3/2}$ \protect\cite{Ernst70}. The
results for $\mathrm{Pe} \leq 1$ are indistinguishable. (b) Time is scaled with
the Stokes time $t_S= a/v_s$ and the velocities are scaled with
$v_s^2$ to highlight hydrodynamic velocity fluctuations.  The straight
line is a fit demonstrating the exponential decay of non-equilibrium
hydrodynamic fluctuations.}
\label{fig3}
\vspace*{-0.5cm}
\end{figure}

Fig.~\ref{fig3} shows the temporal correlation functions along the
direction of sedimentation. At short times the behavior is well
described by exponential Brownian relaxation\cite{Russ89}: $C_{short}(t)$$=$$
\Delta v_{T}^2 \exp\left( -t/ \tau_{B} \right)$.  At intermediate
times it follows the well known algebraic long time tail $C_{long}(t)$$=$$Bt^{-3/2}$, associated with the fact that momentum fluctuations
diffuse away at a finite rate determined by the kinematic viscosity
$\nu$.  Analytical \textit{equilibrium} calculations of
$B$\cite{Ernst70} exactly fit the low Pe \textit{nonequilibrium}
results in Fig.~\ref{fig3}(a) with no adjustable parameters!

\begin{figure}[t]
  \scalebox{0.40}{\includegraphics{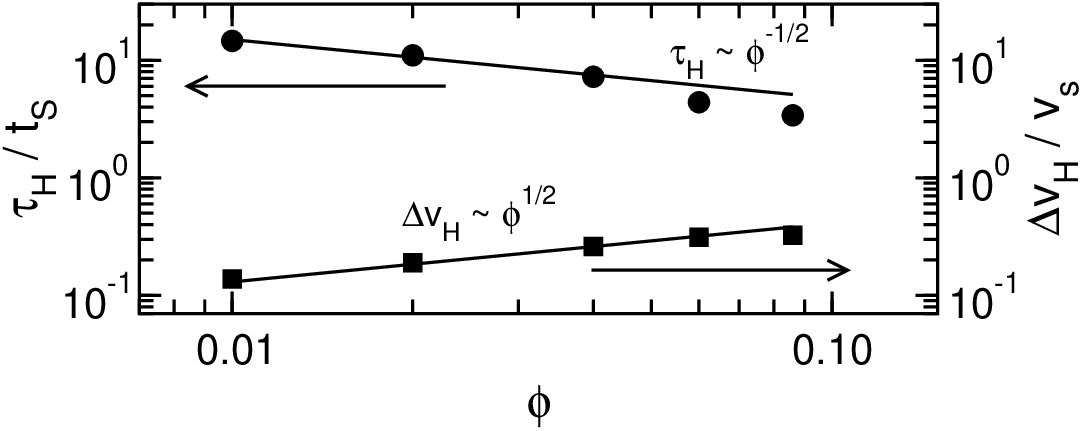}}
\caption{Scaling of the hydrodynamic relaxation times (left scale) and
 velocity fluctuation amplitudes (right scale) with volume
 fraction. Straight lines are expected scalings for an unscreened
 system\protect\cite{CaflishLuke,Hinch}. 
\label{fig4}
}\vspace*{-0.5cm}
\end{figure}

Several experimental studies\cite{Nico95} on the sedimentation of
non-Brownian ($\mathrm{Pe}
\rightarrow \infty$) particles have found an exponential relaxation
of the form
\begin{equation}
C_z(t) = \Delta v_H^2 \exp\left( -t/ \tau_H \right) \mbox{.}
\label{Chydro}
\end{equation}
This non-equilibrium hydrodynamic effect takes place over much longer
time-scales than the initial exponential Brownian relaxation.  The
double-logarithmic Fig.~\ref{fig3}(a) shows that a new mode of
fluctuations becomes distinguishable in our simulations for $\mathrm{Pe} >1$.
In Fig.~\ref{fig3}(b) the correlation functions are scaled with
$v_s^2$ to highlight the nonequilibrium hydrodynamic fluctuations.
For $\mathrm{Pe} \geq 10$ the fluctuations scale onto a single exponential
master curve, similar to the high-Pe experiments\cite{Nico95},
whereas for lower Pe deviations are seen.  From the exponential fit
to Eq.~(\ref{Chydro}), we can estimate the relaxation time $\tau_H$
and the amplitude $\Delta v_H^2$ of the hydrodynamic fluctuations.
These are shown in Fig.~\ref{fig4} for different volume fractions
$\phi$.  The scalings of the relaxation time and fluctuation amplitude
with $\phi$ are consistent with $\Delta v_H^2 \approx v_s^2 L \phi /
(\frac43 \pi a)$ and $\tau_H^2 \approx L^2 / \Delta v_H^2 \approx
t_S^2
\frac43 \pi L / (\phi a)$,  predicted for unscreened
hydrodynamic fluctuations by a simple heuristic argument\cite{Hinch}
akin to that  used by
Caflish and Luke~\cite{CaflishLuke}.

As seen in Fig.~\ref{fig3}, the short time velocity fluctuations are
dominated by thermal fluctuations at all Peclet numbers studied.
By comparing $\Delta v_H$ with $\Delta v_T$, we estimate the
critical $\mathrm{Pe}^*$, above which hydrodynamic fluctuations are 
larger than thermal fluctuations for all $t$~\cite{screening}:
\begin{equation}
\frac{\Delta v_H}{\Delta v_{T}} \approx \frac{(k_BTL\phi
\rho_c)^\frac12}{\gamma} \mathrm{Pe} \equiv \frac{\mathrm{Pe}}{\mathrm{Pe}^*} \mbox{.}
\label{ratio}
\end{equation}
For example, for polystyrene colloids in water ($\eta = 10^{-3}$ Pa s,
$T = 300$ K, $\rho_c = 1050$ kg m$^{-3}$), $\mathrm{Pe}^{*} \approx
\left[(a/10^{-14}\mathrm{m}) / (\phi L/a)\right]^{\frac12} $. For
$\phi = 0.001$, $a = 10^{-6}$ m, and $L/a = 100$ (smaller than the
screening length at this concentration \cite{Nico95,screening}), we find a large value: $\mathrm{Pe}^*
\approx 3 \times 10^4$.

In conclusion, we have adapted a mesoscopic simulation method,
SRD\cite{Male99}, to study sedimentation at finite Peclet numbers.
Hydrodynamic backflow corrections reduce the average sedimentation
velocity $v_s$, irrespective of Pe. Thus,  even when HI are relatively
small, Brownian dynamics simulation methods\cite{Russ89} will yield
qualitatively incorrect results for this problem. Long-time
nonequilibrium velocity fluctuations become evident for $\mathrm{Pe} >1$, and
scale like those for $\mathrm{Pe}\rightarrow\infty$, while short time 
fluctuations are dominated by Brownian forces up to surprisingly large
Pe.  In other words,  neither hydrodynamic
interactions nor Brownian forces can be ignored for a significant parameter regime.

We thank R. Bruinsma, J. Dzubiella, J. Lister and  S. Ramaswamy  for 
helpful
discussions. JTP thanks the EPSRC and IMPACT Faraday, and 
AAL thanks the Royal Society  (London) for financial support.

\FloatBarrier


\begin{thebibliography}{99}
\bibitem{noslip}
Note that the SRD particles interact radially with the colloids,
corresponding to slip boundary conditions. We don't expect any 
qualitative  changes if we were to use no-slip boundary
conditions.
\bibitem{screening}
When hydrodynamic fluctuations are screened\protect\cite{Nico95},
 then $L$ in Eq.~(\protect\ref{ratio}) should be replaced by the 
screening length $\xi(a,\phi)$, which increases $\mathrm{Pe}^*$.
\end{thebibliography}
\end{document}